\documentclass[onecolumn,showpacs]{revtex4}
\usepackage{epsfig,graphicx}
\usepackage{dcolumn}
\usepackage{bm}
\usepackage{ams}
\begin{document}

\title{Temperature dependence of spin depolarization of drifting electrons in n-type
GaAs bulks%
}

\author{Stefano Spezia\footnote{Email: stefano.spezia@gmail.com}, Dominique Persano Adorno, Nicola Pizzolato, Bernardo Spagnolo}
 \affiliation{Dipartimento di Fisica e Tecnologie Relative, \\
Universit\`a di Palermo and CNISM-INFM, \\
Viale delle Scienze, edificio 18, I-90128 Palermo, Italy}

\begin{abstract}
The influence of temperature and transport conditions on the
electron spin relaxation in lightly doped n-type GaAs semiconductors
is investigated. A Monte Carlo approach is used to simulate electron
transport, including the evolution of spin polarization and
relaxation, by taking into account intravalley and intervalley
scattering phenomena of the hot electrons in the medium. Spin
relaxation lengths and times are computed through the
D'yakonov-Perel process, which is the more relevant spin relaxation
mechanism in the regime of interest ($10<T<300$ K). The decay of the
initial spin polarization of the conduction electrons is calculated
as a function of the distance in the presence of a static electric
field varying in the range $0.1- 2$ kV/cm. We find that the electron
spin depolarization lengths and times have a nonmonotonic dependence
on both the lattice temperature and the electric field amplitude.
\end{abstract}


\maketitle

\section{Introduction}
\indent The processing of a high volume of information and world
wide communication are, at the present, based on semiconductor
technology, whereas information storage devices rely on multilayers
of magnetic metals and insulators. Semiconductor spintronics offers
a possible direction towards the development of hybrid devices that
could perform logic operations, communication and storage, within
the same material technology: electron spin could be used to store
information, which could be transferred as attached to mobile
carriers and finally
detected~\cite{FabianDasSarma1999}-\cite{Gul09}.\\
\indent Although these important potential advantages, the designers
of spin devices have to worry about the loss of spin polarization
(spin coherence) before, during and after the necessary
manipulations. In particular, efficient injection, transport,
control and detection of spin polarization must be carefully
treated~\cite{Wolf2001}. Electron-spin states depolarize by
scattering with imperfections or elementary excitations such as
phonons. Hence, for the operability of prospective spintronics
devices, the features of spin relaxation at relatively high
temperature jointly with the influence of transport conditions
should be firstly understood
~\cite{FabianDasSarma1999,Barry2003}.\\
\indent In recent years there was a proliferation of experimental
works in which the influence of transport conditions on relaxation
of spins in semiconductors has been
investigated~\cite{Kikkawa98}-\cite{Furis2006}. Even though for high
speed transfer of information, high external electric fields must be
used, up now only the influence of low electric fields ($F<0.1$
kV/cm) on coherent spin transport has been investigated and very
little is known about the effects of higher
electric fields~\cite{Sanada2002} or high lattice temperatures.\\
\indent Despite of this great experimental interest, few theoretical
works ~\cite{Perel1971,Hruska2006} and simulative
studies~\cite{Barry2003,Saikin2003,Saikin2005}, have been carried
out. Theoretical approaches to describe spin dynamics and
spin-polarized electron transport include the two-component
drift-diffusion model~\cite{Hruska2006}, Monte Carlo techniques to
solve the Boltzmann equation~\cite{Barry2003} and microscopic
approaches solving the Bloch kinetic equations~\cite{Weng2004}.
However, a comprehensive theoretical investigation of the influence
of transport conditions on the spin depolarization in semiconductor
bulk structures, in a wide range of values of temperature and
amplitude of external fields, is lacking.\\
\indent Inducing spin polarization in a semiconductor, such as GaAs
and Si, can be done efficiently and at reasonable current levels by
electrical transfer of spins from a ferromagnetic metal across a
thin tunnel barrier, at low temperatures ($5-150$
K)~\cite{Lou07,Jon07}. Very recently, electrical injection of spin
polarization in n-type and p-type silicon at room-temeperature have
been experimentally carried out~\cite{Dash2009}. These promising
experimental results for development of spintronics devices suggest
that it is important investigate the spin coherence up to room
temperature. Earlier Monte Carlo simulation has revealed that the
presence of an external electric field can accentuate spin
relaxation in GaAs bulk materials~\cite{Barry2003}. In this work,
solving the transport and spin dynamics stochastic differential
equations by a semiclassical Monte Carlo approach, we estimate the
spin lifetimes and depolarization lengths of an ensemble of
electrons, for intermediate values of the electric field ($0.1-2$
kV/cm) and lattice temperatures in the range $10<T<300$
K.\\
\indent The paper is organized as follows: in Sec.~2 the
semiconductor physical model and the Monte Carlo approach are
presented; in Sec.~3 the numerical results are given and discussed.
Final comments and conclusions are given in Sec.~4.

\section{Semiconductor model and spin-polarized electron transport calculation}
\label{Model}

\subsection{Semiconductor model and semiclassical Monte Carlo approach}
\label{MCP}

\indent The study of the transport properties and the spin
relaxation process in a semiconductor in the presence of an external
field is not simple, especially when the field is very strong. In
this case, it is preferable to perform a numerical simulation of the
process. The Monte Carlo method presents the remarkable advantage of
giving a detailed description of the particles motion in the
semiconductor taking into account the scattering mechanisms, and
allows us to obtain all the needed information, such as average
velocity of the electrons, temperature, current density, etc.,
directly without the need of calculating the electron distribution
function. The time of free flight (time interval between two
collisions), the collisional mechanisms, the scattering angle, and
all the parameters of the problem are chosen in a stochastic way,
making a mapping between the probability density of the given
microscopic process and a uniform distribution of random numbers. \\
\indent The Monte Carlo algorithm, developed for simulating the
motion of electrons in a GaAs semiconductor, follows the standard
procedure described in Ref.~\cite{Persano}. Here, we incorporate the
description of the electron spin dynamics by following a standard
semiclassical formalism. We assume that the spatial electron
transport is well described by the Boltzmann equation and that the
electrons move along classical trajectories between two scattering
events. The conduction bands of GaAs are the $\Gamma$-valley, four
equivalent L-valleys and three equivalent X-valleys. The parameters
of the band structure and scattering mechanisms are taken from
Ref.~\cite{Persano}. In this work Monte Carlo simulations of
electron transport and spin depolarization dynamics are limited to
low-energy regime with the electric field amplitude varying in the
range $0.1-2~$kV/cm. In this energy range the electrons can be found
only in the $\Gamma$-valley. Our computations include the effects of
the nonparabolicity of the band structure and, among many different
scattering mechanism, electron scattering due to ionized impurities,
acoustic, piezoelectric and polar optical phonons in the
$\Gamma$-valley. The scattering probabilities are calculated by
 the Fermi Golden Rule and the scattering events are considered
 instantaneous. We assume field-independent scattering
probabilities; accordingly, the influence of the external fields is
only indirect through the field-modified electron velocities.
Nonlinear interactions of the field with the lattice and bound
carriers is neglected. We neglect also electron-electron
interactions and consider electrons to be
independent~\cite{Kiselev2000}. All simulations are performed in a
n-type GaAs bulk with a free electrons concentration $n = 10^{13}
cm^{-3}$. We have assumed that all donors are ionized and that the
free electron concentration is equal to the doping concentration.

\subsection{ Spin relaxation dynamics}
\label{Calculation1} \indent The spin depolarization of drifting
electrons is analyzed for a lattice temperature $T$ varying in the
range $10 < T < 300$ K. For these values of $T$ the D'yakonov-Perel
process is the more relevant spin relaxation
mechanism~\cite{Furis2006}. This mechanism, effective in the
intervals between collisions, is related to the spin-orbit splitting
of the conduction band in non-centrosymmetric semiconductors like
GaAs~\cite{DyakonovEd,Perel1971,Dyakonov2006} .

In a semiclassical formalism the effective single-electron
Hamiltonian which accounts for the spin-orbit interaction term is
\begin{equation}
H = H_{0} + H_{SO} \label{Hamiltonian}
\end{equation}
where $H_{0}$ is the self-consistent electron Hamiltonian in the
Hartree approximation, including also interactions with impurities
and phonons. The spin-dependent term $H_{SO}$ may be written as
\begin{equation}
H_{SO} = \frac{\hbar}{2}\vec{\sigma}\cdot\vec{\Omega}_{eff},
\label{HamiltonianSO}
\end{equation}
and can be viewed as the energy of a spin in an effective magnetic
field that causes electron spin to precess. $\vec{\Omega}_{eff}$ is
a vector depending on the orientation of the electron momentum
vector with respect to the crystal axes (xyz). Near the bottom of
the $\Gamma$-valley, the effective magnetic field can be written as
~\cite{Ivchenko1997}
\begin{equation}
\vec{\Omega}_{eff} =
\beta_{\Gamma}[k_{x}(k_{y}^{2}-k_{z}^{2})\hat{x}+k_{y}
(k_{z}^{2}-k_{x}^{2})\hat{y}+k_{z}(k_{x}^{2}-k_{y}^{2})\hat{z}]
\label{effectivefieldgammavalley}
\end{equation}
where $k_{i}$ are the components of the electron wave vector, and
$\beta_{\Gamma}$ the spin-orbit coupling
coefficient~\cite{Perel1971}. In particular,
\begin{equation}
\beta_{\Gamma}=\frac{\alpha\hbar^{2}}{m\sqrt{2m
E_{g}}}(1-\frac{E(\vec{k})}{E_{g}}\frac{9-7\eta+2\eta^{2}}{3-\eta})
\label{betaGammavalley}
\end{equation}
where $\alpha=0.029$ is a dimensionless material-specific parameter
which gives the magnitude of the spin-orbit splitting, $\eta =
\Delta/(E_{g}+\Delta)$, with $\Delta=0.341$ eV the spin-orbit
splitting of the valence band, $E_{g}$ is the energy separation
between the conduction band and valence band at the $\Gamma$ point
and $m$ is the effective mass. In Eq.~(\ref{betaGammavalley}), we
consider the effects of nonparabolicity on the spin-orbit splitting
in $\Gamma$-valley, estimated by Pikus and Titkov~\cite{Pikus89}.

The quantum-mechanical description of the evolution of the spin
$1/2$ is equivalent to the evolution of the classical momentum
$\vec{S}$ under an effective magnetic field $\vec{\Omega}_{eff}$
with the equation of motion
\begin{equation}
\frac{d\vec{S}}{dt}=\vec{\Omega}_{eff}\times\vec{S}. \label{Poisson}
\end{equation}
In Eq.~(\ref{Poisson}), the scattering reorients the direction of
the precession axis, making the orientation of the effective
magnetic field random and trajectory-dependent, thus leading to spin
relaxation (dephasing)~\cite{Perel1971}. The reciprocal effect of
the electron spin evolution on the orbital motion through spin-orbit
coupling can be ignored due to the large electron kinetic energy in
comparison with the typical spin splittings and strong change of the
momentum in scattering events~\cite{Kiselev2000}. In this
modelization the  scattering processes are considered
spin-independent.

\subsection{Calculation of spin depolarization times and lengths}
\label{Calculation2}
\begin{figure}[htbp]
\centering
\resizebox{0.80\columnwidth}{!}{%
\includegraphics*[height=7cm,width=11cm]{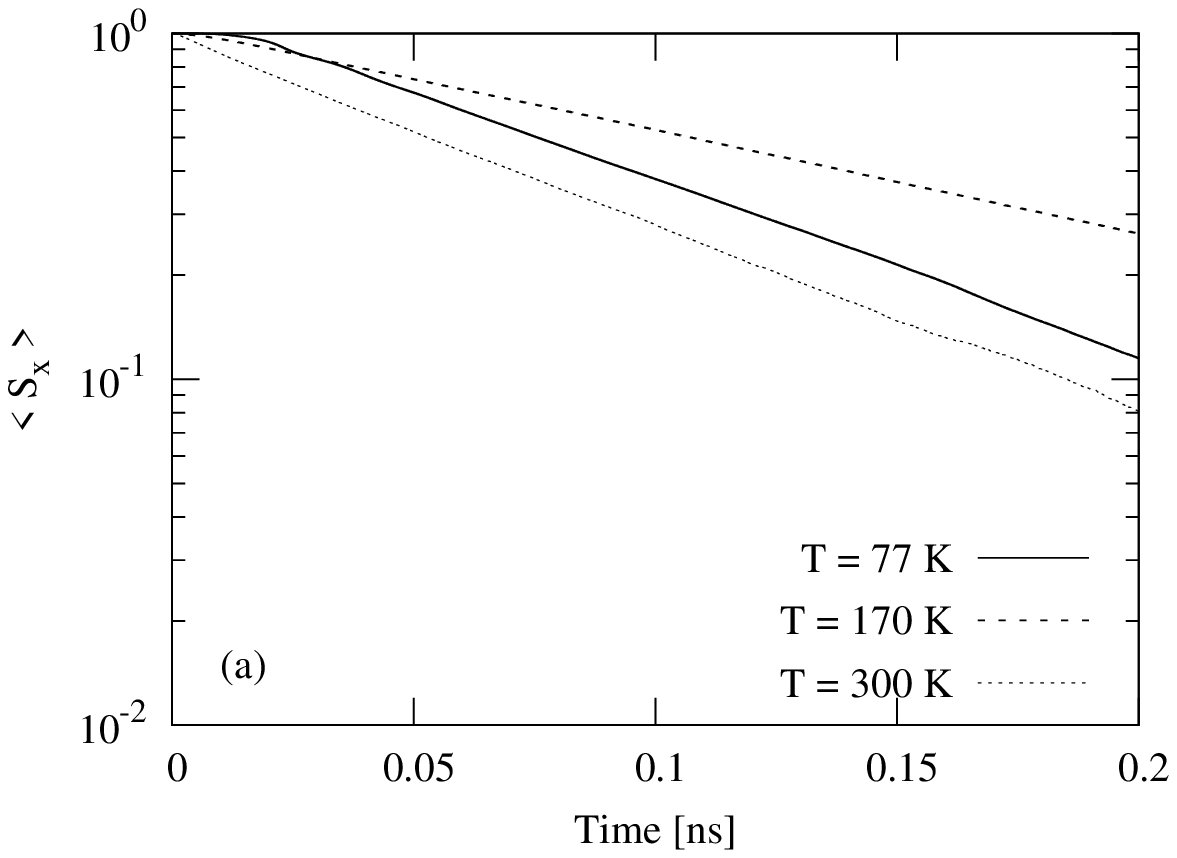}
}
\resizebox{0.80\columnwidth}{!}{%
\includegraphics*[height=7.42cm,width=11cm]{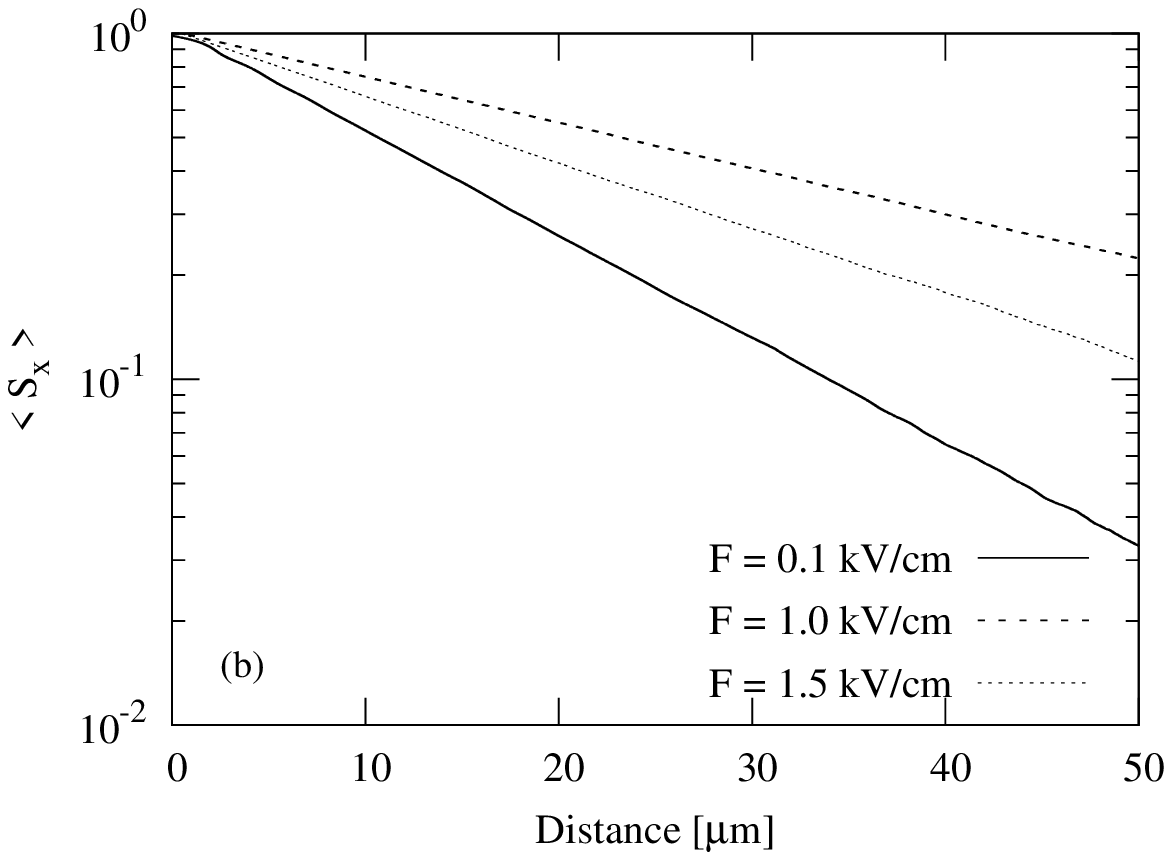}
} \caption{(a) Spin polarization $\langle S_{x}\rangle$ as a
function of time, with field amplitude $F=0.1$ kV/cm, at three
different values of temperature; (b) Spin polarization $\langle
S_{x}\rangle$ as a function of distance at $T=77$ K, for three
different values of the electric field amplitude.} \label{Sx}
\end{figure}
\indent The dependence of spin relaxation times and lengths on
temperature and driving electric field has been investigated by
simulating the dynamics of $5\cdot10^{4}$ electrons, initially
polarized $(\langle \vec{S}\rangle=1)$ along the $\hat{x}$-axis at
the injection plane $(x_{0}=0)$. We calculate $\langle
\vec{S}\rangle$ as a function of time by averaging over the ensemble
of electrons. In Fig.~\ref{Sx} (a), we show the electron average
polarization $\langle S_{x}\rangle$, calculated as a function of
time in the presence of an electric field, with amplitude $F=0.1$
kV/cm and directed along $\hat{x}$-axis, for three different values
of temperature. In Fig.~\ref{Sx} (b), we show the same component of
spin polarization $\langle S_{x}\rangle$, calculated at $T=77$ K, as
a function of the distance traveled by the center of mass of the
electron cloud from the injection plane, for three different values
of the external field amplitude. Since $\langle S_{x}\rangle$ is
found to decrease with both time and distance by showing an almost
linear trend in a semi-log plot, the spin relaxation times $\tau$
and lengths $L$ are estimated by considering the spin depolarization
as an exponentially process dependent on time and
distance~\cite{Barry2003}. If $\langle S_{x}\rangle$ and $\langle x
\rangle$ are the mean polarization along $\hat{x}$-axis and the mean
position of the ensemble of the electrons as a function of time,
$\tau$ and $L$ are chosen to be characteristic time and distance
such that
\begin{equation}
\langle S_{x}\rangle = A\cdot exp(-t/\tau)= B\cdot exp(-\langle x
\rangle/L), \label{exponential}
\end{equation}
with $A$ and $B$ normalization factors. $L$ and $\tau$ satisfy the
relation $L= v_d\cdot\tau$, where $v_d$ is the average drift
velocity.

\section{Numerical results and discussion}
\label{Results}
\begin{figure}[htbp]
\centering
\resizebox{0.80\columnwidth}{!}{%
\includegraphics*[height=7cm,width=11cm]{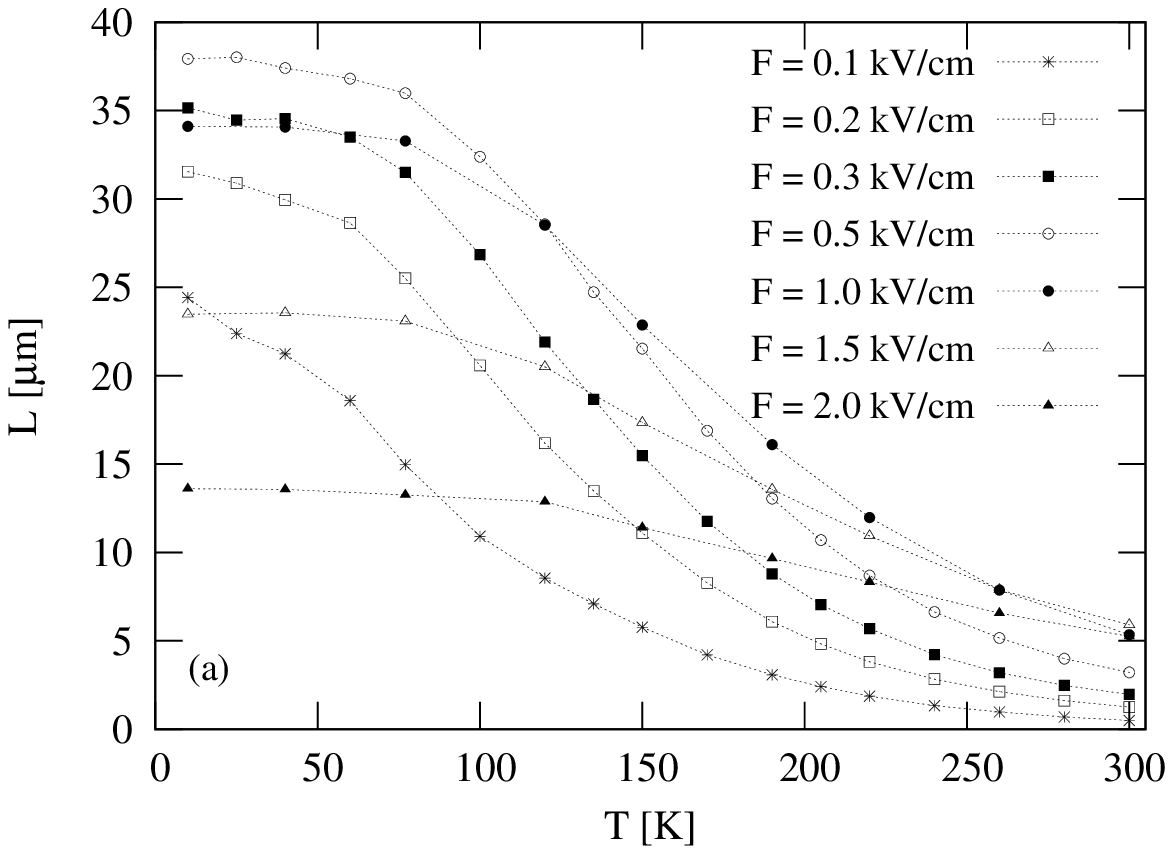}}
\resizebox{0.80\columnwidth}{!}{%
\includegraphics*[height=7.42cm,width=11cm]{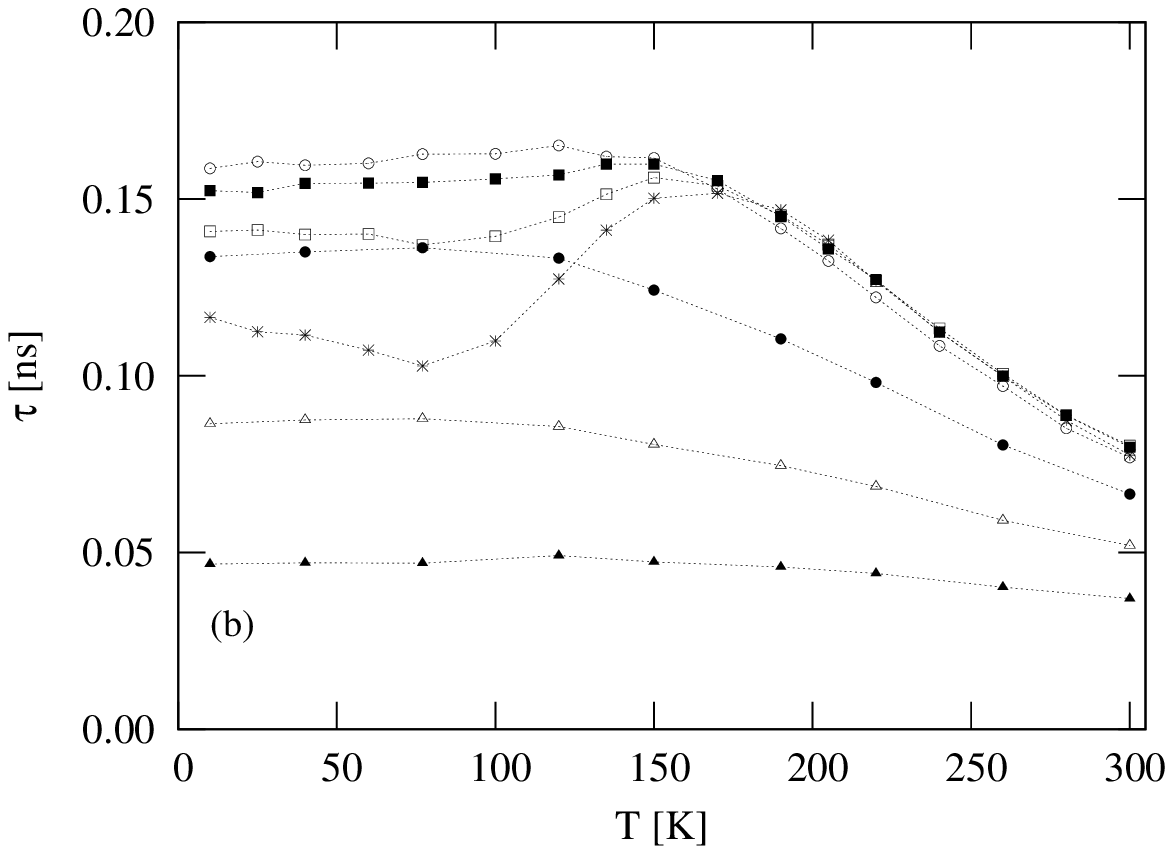}
} \caption{(a) Spin depolarization length $L$ and (b) spin
depolarization time $\tau$ as a function of the temperature $T$,
plotted for several values of the electric field amplitude $F$.}
\label{vsT}
\end{figure}
\indent In Fig.~\ref{vsT} we show the spin electron relaxation
length $L$ [panel (a)] and the spin depolarization time $\tau$
[panel (b)] as a function of the lattice temperature, for different
values of the electric field amplitude, namely $F = 0.1, 0.2, 0.3,
0.5, 1.0, 1.5, 2.0$ kV/cm. For a fixed electric field, the spin
electron relaxation length is a monotonic decreasing function of the
temperature. When $F=0.5$ kV/cm, $L$ shows its maximum value,
remaining greater than $35~\mu$m up to $T\simeq80$ K. Furthermore,
for field amplitudes greater than $1$ kV/cm, the spin depolarization
length remains almost constant for $T<100$ K. At room temperature
the maximum value of $L$ ($\sim6~\mu$m) is obtained for
F$\geq1$ kV/cm. \\
\indent The relaxation time $\tau$ shows, instead, a nonmonotonic
behavior with the temperature [see Fig.~\ref{vsT} (b)]. In
particular, the curves obtained with $F=0.1$ and $0.2$ kV/cm exhibit
a minimum at $T\sim80$ K and an increase in the range $100-170$ K.
For temperatures greater than $170$ K, all curves with a field
amplitude up to $0.5$ kV/cm show a common decreasing trend. The
longest value of spin coherence time is achieved for the field
amplitude $F=0.5$ kV/cm for almost the entire range of temperatures.
For higher values of $F$, the spin depolarization time strongly
decreases, becoming
nearly temperature-independent for $F>1.5$ kV/cm.\\
\indent As the temperature increases, the scattering rate increases
too, and hence the ensemble of spins loses its spatial order faster,
resulting in a faster spin relaxation. This temperature dependence
becomes less evident at higher amplitudes of the driving electric
field, where, because of the greater drift velocities, the
polarization loss is mainly due the to strong effective magnetic
field. At very low electric fields, the spin dephasing is, instead,
primarily caused by the multiple scattering events. The
nonmonotonicity of $\tau$ can be ascribed by the progressive change,
with the increase of the temperature, of the dominant scattering
mechanism from acoustic phonons and ionized impurities to polar
optical phonons \cite{Dzhioev2004}. Following the standard theory of
D'yakonov-Perel, $\tau^{-1}$ is proportional to the third power of
the temperature $T$ and linearly depends on the momentum relaxation
time $\tau_p$ \cite{Perel1971}. An increase of the temperature
initially leads to a slightly decrease of $\tau$; for temperatures
greater than $\sim100$ K the electrons start to experience
scattering by polar optical phonons. This switching on leads to an
abrupt decrease of $\tau_p$ that, for lattice temperatures in the
range $100-150$ K, results  more effective than the increase of $T$,
giving rise to the observed increase of $\tau$. For temperatures
greater than $150$ K this latter effect is no more relevant.\\
\begin{figure}[htbp]
\centering
\resizebox{0.80\columnwidth}{!}{%
\includegraphics*[height=7cm,width=11cm]{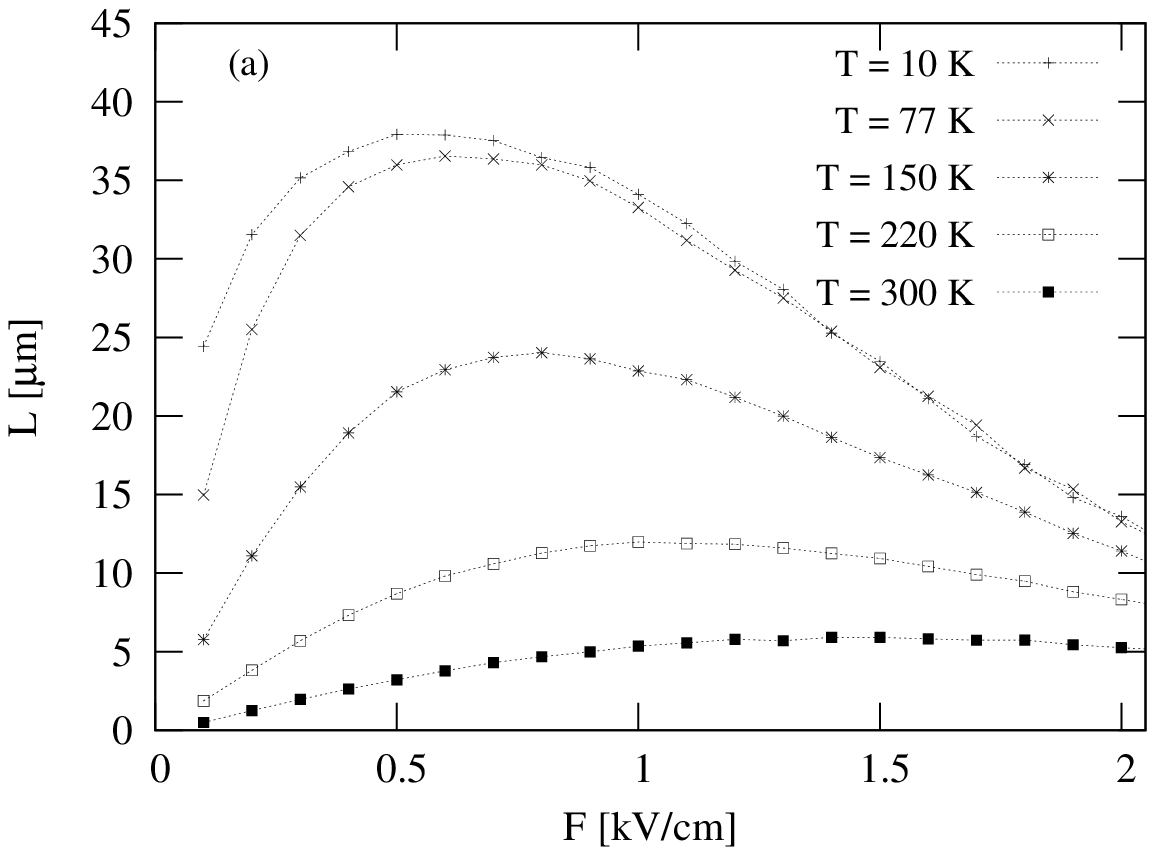}}
\resizebox{0.80\columnwidth}{!}{%
\includegraphics*[height=7.42cm,width=11cm]{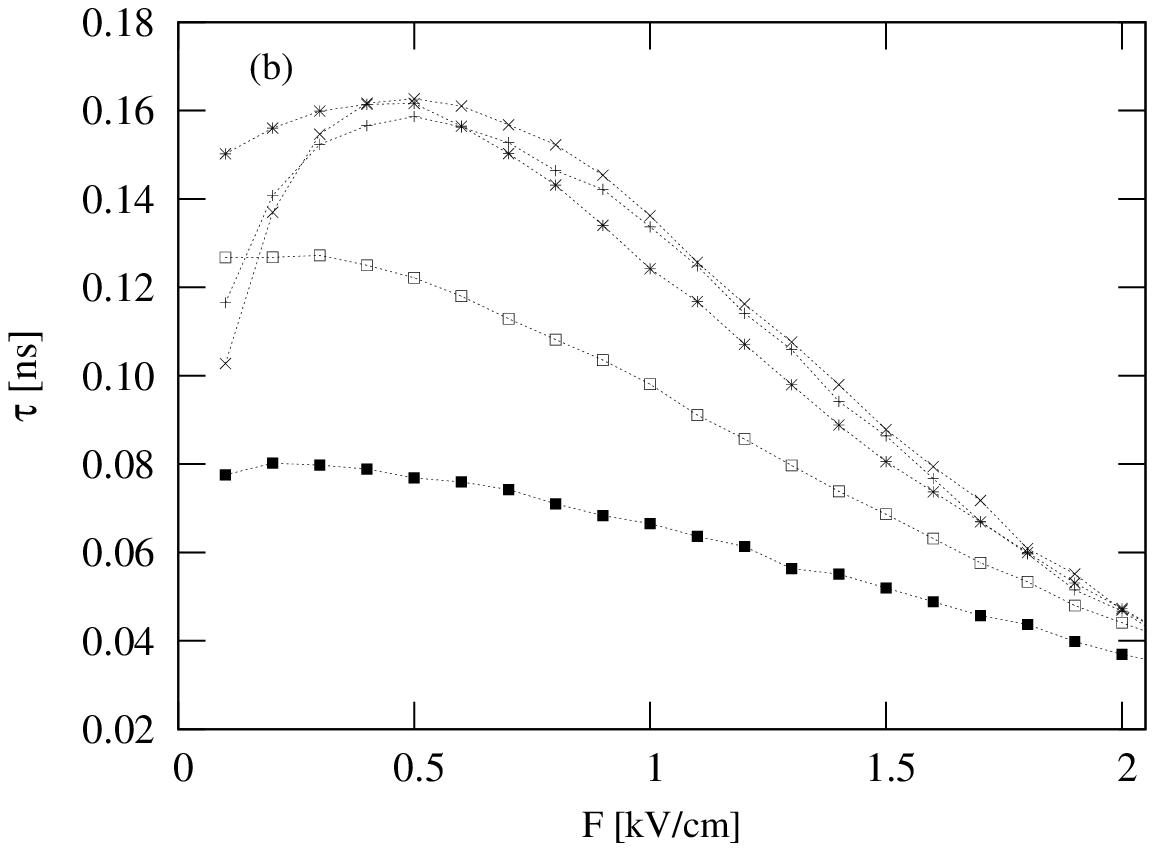}
} \caption{(a) Spin depolarization length $L$ and (b) Spin
depolarization time $\tau$ as a function of the electric field
amplitude $F$, plotted for several values of the temperature $T$.}
\label{vsF}
\end{figure}
\indent In Fig.~\ref{vsF} we plot the spin depolarization length $L$
[panel (a)] and the spin depolarization time $\tau$ [panel (b)] as a
function of the electric field amplitude, for different values of
the lattice temperature. The spin relaxation lengths show a marked
maximum that rapidly reduces its intensity, widens and moves towards
higher electric field amplitudes with the increasing of the
temperature. For temperatures $T\leq150$ K the decoherence times
plotted in Fig.~\ref{vsF} (b) show a nonmonotonic behavior. For
$F>0.5$ kV/cm, $\tau$ lightly depends on the temperature up to
$T\sim150$ K. At higher temperatures, the spin electron relaxation
time becomes a monotonic decreasing function of the electric field
intensity.\\
\indent The presence of maxima in the spin depolarization length at
intermediate fields can be explained by the interplay between two
competing factors: in the linear regime, as the field becomes
larger, the electron momentum and the drift velocity increase in the
direction of the field. On the other hand, the increased electron
momentum also brings about a stronger effective magnetic field, as
shown in Eq.~\ref{effectivefieldgammavalley} ~\cite{Barry2003}.
Consequently, the electron precession frequency becomes higher,
resulting in faster spin relaxation (i.e., shorter spin relaxation
time). For $F<0.5$ kV/cm and $T\leq150$ K the nonmonotonic behavior
of the relaxation time reflects the complex scenario described
above, caused by the triggering of scattering mechanisms having
different rates of occurrence.

\section{Conclusions}
\indent For the extensive utilization of spintronics devices, the
features of spin decoherence at relatively high temperature, jointly
with the influence of transport conditions, should be fully
understood. In this work, by using a semiclassical Monte Carlo
approach, we have estimated the spin mean lifetimes and
depolarization lengths of an ensemble of conduction electrons in
lightly doped n-type GaAs crystals, in a wide range of lattice
temperatures ($10<T<300$ K) and field amplitudes ($0.1<F<2$ kV/cm).
We have shown that, under particular conditions, also at
temperatures greater than the liquid-helium temperature, it is
possible to obtain very long spin relaxation times and relaxation
lengths. These are essential for the high performance of spin-based
devices, in order to extend the functionality of conventional
devices to higher working temperatures and higher electric field
amplitudes and to allow the development of new information
processing systems. In particular, for $F=0.5$ kV/cm we achieve the
longer value of spin lifetime ($\tau>0.15$ ns) up to a temperature
$T=150$ K. At room temperatures, we obtain a coherence
length of about $6~\mu$m, nearly independent from the intensity of the electric field.\\
\indent Furthermore, depending on the interplay between the external
electric field and the different collisional mechanisms with
increasing electron energy, we find very interesting nonmonotonic
behavior of spin lifetimes and depolarization lengths as a function
of temperature and electric field amplitude. This point deserves
further investigations.

Understanding these phenomena could lead to high temperature and
high field improvement of the gating mechanisms engineering
spin-based devices.

\section*{Acknowledgments}
This work was partially supported by MIUR and CNISM-INFM.

\end{document}